\documentclass[pra,twocolumn, aps, preprintnumbers,superscriptaddress]{revtex4-2}
\usepackage[colorlinks=true,citecolor=blue,urlcolor=black]{hyperref}
\usepackage{verbatim}
\usepackage{amsmath}
\usepackage{latexsym}
\usepackage{revsymb}
\usepackage{yfonts}
\usepackage{ifthen}

\usepackage{natbib}
\usepackage{amsfonts}
\usepackage{amsmath}
\usepackage{amssymb}
\usepackage{amsthm}
\usepackage{graphicx}
\usepackage{bm}
\usepackage{bbm}
\usepackage{epsfig,color,amssymb}
\usepackage{subfigure}
\usepackage{amsfonts}
\usepackage{amscd}
\usepackage{amsmath}
\usepackage{multirow}
\usepackage{chemarrow}
\usepackage{dcolumn}
\usepackage{bm}
\usepackage{graphicx}
\usepackage{enumerate}
\usepackage{epsfig}
\usepackage{subfigure}
\usepackage{xcolor}
\usepackage{multirow}
\usepackage{ulem}
\usepackage{braket}
\usepackage{comment}
\usepackage{enumitem}
\usepackage{amsthm}
\renewcommand{\emph}[1]{\textit{#1}}

\begin{document}
	
	\title{Differential Phase Shift Quantum Secret Sharing Using a Twin Field with Asymmetric Source Intensities}

	\author{Zhao-Ying Jia}
	\author{Jie Gu}
	\author{Bing-Hong Li}
	\author{Hua-Lei Yin}\email{hlyin@nju.edu.cn}
	\author{Zeng-Bing Chen}\email{zbchen@nju.edu.cn}
	\affiliation{National Laboratory of Solid State Microstructures, School of Physics, and Collaborative Innovation Center of Advanced Microstructures, Nanjing University, Nanjing 210093, China}
	
	
	\begin{abstract}
	As an essential application of quantum mechanics in classical cryptography, quantum secret sharing has become an indispensable component of quantum internet. Recently, a differential phase shift quantum secret sharing protocol using a twin field has been proposed to break the linear rate-distance boundary. However, this original protocol has a poor performance over channels with asymmetric transmittances. To make it more practical, we present a differential phase shift quantum secret sharing protocol with asymmetric source intensities and give the security proof of our protocol against individual attacks. Taking finite-key effects into account, our asymmetric protocol can theoretically obtain the key rate two orders of magnitude higher than that of the original protocol when the difference in length between Alice’s channel and Bob’s is fixed at 14 km. Moreover, our protocol can provide a high key rate even when the difference is quite large and has great robustness against finite-key effects. Therefore, our work is meaningful for the real-life applications of quantum secret sharing.
	\end{abstract}

	\maketitle
	\section{INTRODUCTION}
	Secret sharing is a cryptographic protocol in which a dealer splits a secret into several parts and distributes them among various players. The secret can be recovered only when a sufficient number of players (authorized subsets) cooperate to share their parts of the secret. The classical secret sharing scheme was first introduced independently by Shamir \cite{shamir1979share} and Blakley \cite{Blakley} in 1979, followed by plenty of variations \cite{brickell1989some}.
However, all existing classical secret sharing schemes are not perfectly secure from eavesdropping attacks~\cite{365700}. 

As the combination of classical secret sharing and quantum mechanics, quantum secret sharing (QSS) is more secure due to the excellent properties of quantum theory and has become one of the most attractive research topics in the quantum cryptography. In 1999, Hillery et al. \cite{hillery1999quantum} firstly proposed a protocol of QSS using a three-photon Greenberger--Horne--Zeilinger (GHZ) state. Afterwards, this protocol was generalized into an arbitrary number of parties based on multi-particle entanglement states \cite{xiao2004efficient}, and later to multi-particle $d$-dimensional entanglement states \cite{yu2008quantum}.
From then on, much
theoretical~\cite{PhysRevA.64.042311,PhysRevA.71.012328,keet2010quantum,fu2015long,PhysRevA.91.022330,PhysRevA.92.030302,PhysRevA.95.012315,9006878} and experimental~\cite{PhysRevA.63.042301,chen2005experimental,PhysRevLett.98.020503,bell2014experimental} attention has focused on QSS using multi-particle entangled states. However, it is a tremendous challenge to prepare a multiparty entanglement state with high fidelity and efficiency, which makes particle entanglement-based QSS unscalable. To circumvent the problems, differential phase shift QSS scheme using coherent light~\cite{Inoue:08}, similar to those used
in {quantum key distribution} (QKD)~\cite{PhysRevLett.89.037902,PhysRevA.68.022317,Honjo:04,Takesue_2005,PhysRevA.73.012344,Diamanti:06}, has been proposed and~implemented.

Nevertheless, the linear rate-distance limitation constricts the key rate and transmission distance of QSS 
~\cite{takeoka2014fundamental,pirandola2017fundamental}. 
Recently, to exceed the linear bound and further enhance the practical performance of QSS, a differential phase shift quantum secret sharing (DPSQSS) protocol~\cite{gu2021differential} using a twin field (TF)~\cite{lucamarini2018overcoming} has been proposed. 
Unfortunately, this protocol suffers from low key rate and short transmission distance over channels with different transmittances, 
which constrains its application in a practical network setting.

Here, we propose an asymmetric differential phase shift quantum secret sharing protocol using twin field~\cite{lucamarini2018overcoming,ma2018phase,PhysRevA.98.062323,lin2018simple,yin2019measurement,PhysRevApplied.11.034053,curty2019simple,yin2019coherent,hu2019sending,grasselli2019asymmetric,maeda2019repeaterless,minder2019experimental,PhysRevLett.123.100505,PhysRevLett.123.100506,zhong2021proof} ideas and give the security proof of this protocol against individual attacks. The key point of our method is that Alice and Bob can adjust their source intensities independently to effectively compensate for channel asymmetry.
The numerical results show that our protocol is robust against finite-key effects~\cite{tomamichel2012tight,Grasselli_2018,yin2020tight,yin2019finite} and can theoretically provide a two orders of magnitude higher key rate than the original protocol with the length difference between Alice’s channel and Bob’s fixed at 14 km. Furthermore, our protocol still obtains a high key rate when the difference in length is fixed at 50,100 km, whereas no keys are obtained with the same difference in the original protocol. Therefore, our work represents a further step along the progress of practical QSS.
	
	\section{TF-DPSQSS Protocol with Asymmetric Source Intensities}
\begin{figure}
		\centering
		\includegraphics[width=8.6cm]{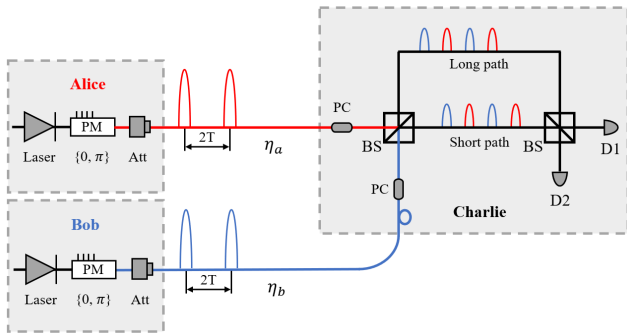}
		\caption{Configuration of our quantum secret sharing protocol. Weak coherent pulse sources (Laser); phase modulator (PM); signal attenuator (Att); transmittance of channels between Alice (Bob) and Charlie ($\eta_a,\eta_b$); polarization control (PC); beam splitter (BS); single photon detector (D1, D2). In Charlie's measurement area, two pulse trains are first polarization-modulated by polarization controls to correct their polarization for interference. Then, Charlie divides each incoming pulse into two paths and recombines them by a 50:50 beam splitter, where the path-length difference is set equal to the time T. Detectors are placed at the two outputs of the recombining beam splitter. At the detectors, the partial wave functions of two senders' pulses that are in the same time slots interfere with each other.
		} \label{fig1}
	\end{figure}

	The schematic diagram of our asymmetric protocol is shown in Fig.~\ref{fig1}, where Alice and Bob have partial keys for deciphering, and Charlie has a full key for ciphering. The two senders, Alice and Bob, independently prepare two trains of weak coherent pulses whose intensities are different and phases are randomly modulated to be 0 or $\pi$. The coherent pulses are sent through the quantum channels and received by the trusted third party, Charlie, who measures them using an unbalanced interferometer. The details of the protocol are shown as follows.
	
	{Preparation:} Alice (Bob) prepares a weak coherent pulse train and phase-modulates each pulse randomly by 0 or $\pi$. The coherent states can be denoted as 
$\left|\psi_a\right\rangle=\underset{n=1}{ \overset{N} {\otimes }}\left|\sqrt{\mu_a} e^{i{\phi_{n}^a}}\right\rangle$
$\left(|\psi_b\right\rangle=\underset{n=1}{ \overset{N} {\otimes }}\left|\sqrt{\mu_b} e^{i{\phi_{n}^b}}\right\rangle)$. Then, she (he) sends out the coherent state pulse whose period is 2T to Charlie with an average photon number less than one per pulse. Alice
(Bob) records her (his) logic bits of each time slot as ``0'' (``1'') when her (his) modulated phase is 0 $(\pi)$. 
We denote the sequence of detection events time slots as {$n$} 
$\in\left\{1,2,3,...,2N\right\}$, where $N$ is the total number of pulses sent by Alice (Bob).
The phase shift $\phi_{n}^a\left(\phi_{n}^b\right)\in\left\{0,\pi\right\}$ is the phase induced by the phase modulator on pulse $n$ ignoring the global phase, and intensities $\mu_a$, $\mu_b$ are corresponding to Alice and Bob, respectively.

{Measurement:} As illustrated in Fig.~\ref{fig1}, while measuring the signal, Charlie records the photon detection time and which detector clicks. When the detection event time slots are corresponding to the time $2kT$, detector 1 will click for 0 phase difference between the two senders' pulses and detector 2 will click for $\pi$ phase difference. When the detection event time slots are corresponding to the time $(2k+1)T$, detector 1 will click for $\pi$ phase difference and detector 2 will click for 0 phase difference. Note that if both detectors click, Charlie randomly chooses one detector click to record. Here, a photon is detected occasionally and randomly because the received signal power is smaller than one photon per pulse.

Using the above setup, Charlie creates his key shown in Table.~\ref{tab1} . ``0'' means that the modulated phases in one time slot imposed by Alice and Bob are $\left\{0, 0\right\}$ or $\left\{\pi, \pi\right\}$, and``1'' means that the modulated phases in one time slot imposed by Alice and Bob are $\left\{0, \pi\right\}$ or $\left\{\pi,0\right\}$. Thus, Charlie's bits are exclusive OR of Alice's and Bob's bits. That is, Alice and Bob know Charlie’s key bits only when they cooperate, and the QSS operation is accomplished.

\begin{table}
	\centering
	\caption{Logic bits held by Charlie corresponding to different detection event time slots when detector 1 or detector 2 clicks.}\label{tab1}
	\begin{tabular}{ccccc}
		\hline
		\hline
		 ~~~~~&Detector$1$ ~~~~~& Detector$2$\\
		\hline
		$2kT$~~~~~& ``0''~~~~~ & ``1''\\
		\hline
		$(2k+1)T$~~~~~& ``1''~~~~~ & ``0''\\
		\hline
		\hline
	\end{tabular}
\end{table}

{Parameter estimation:} Charlie randomly chooses recorded detection times and Alice and Bob alternatively disclose her or their test bit first in the chosen time slots through a public channel. Then, Charlie will get the quantum bit error rate (QBER) and make a decision whether they discard all their bits and restart the whole QSS at Step 1 {(preparation)}.

{Postprocessing:} After calculating the QBER, Alice, Bob, and Charlie will conduct classical error correction
and privacy amplification to distill the final full key and partial~keys.

	\section{Proof of Security}
	In this section, we will discuss the security of our protocol against eavesdropping. Because of the equivalence~\cite{gu2021differential} 
between our asymmetric protocol and differential phase shift QSS, we can apply the conclusion in differential phase shift quantum key distribution~\cite{PhysRevA.73.012344} to the analysis of both an external eavesdropper and an internal eavesdropper in our protocol.

\subsection{External Eavesdropping}

Firstly, Eve cannot obtain full key information by beam-splitting attacks and intercept-resend attacks, which will result in bit errors in the secret key \cite{PhysRevA.68.022317}. As for a general individual attack, based on the assumption that Eve will conduct the same attack in differential phase shift QSS~\cite{Inoue:08}, we can derive that information leakage to Eve is given by a fraction $\mu_a(1-\eta_a)+\mu_b(1-\eta_b)$ of the sifted key \cite{PhysRevA.73.012344}.

\subsection{Internal Eavesdropping}

\begin{figure}
	\includegraphics[width=8.6 cm]{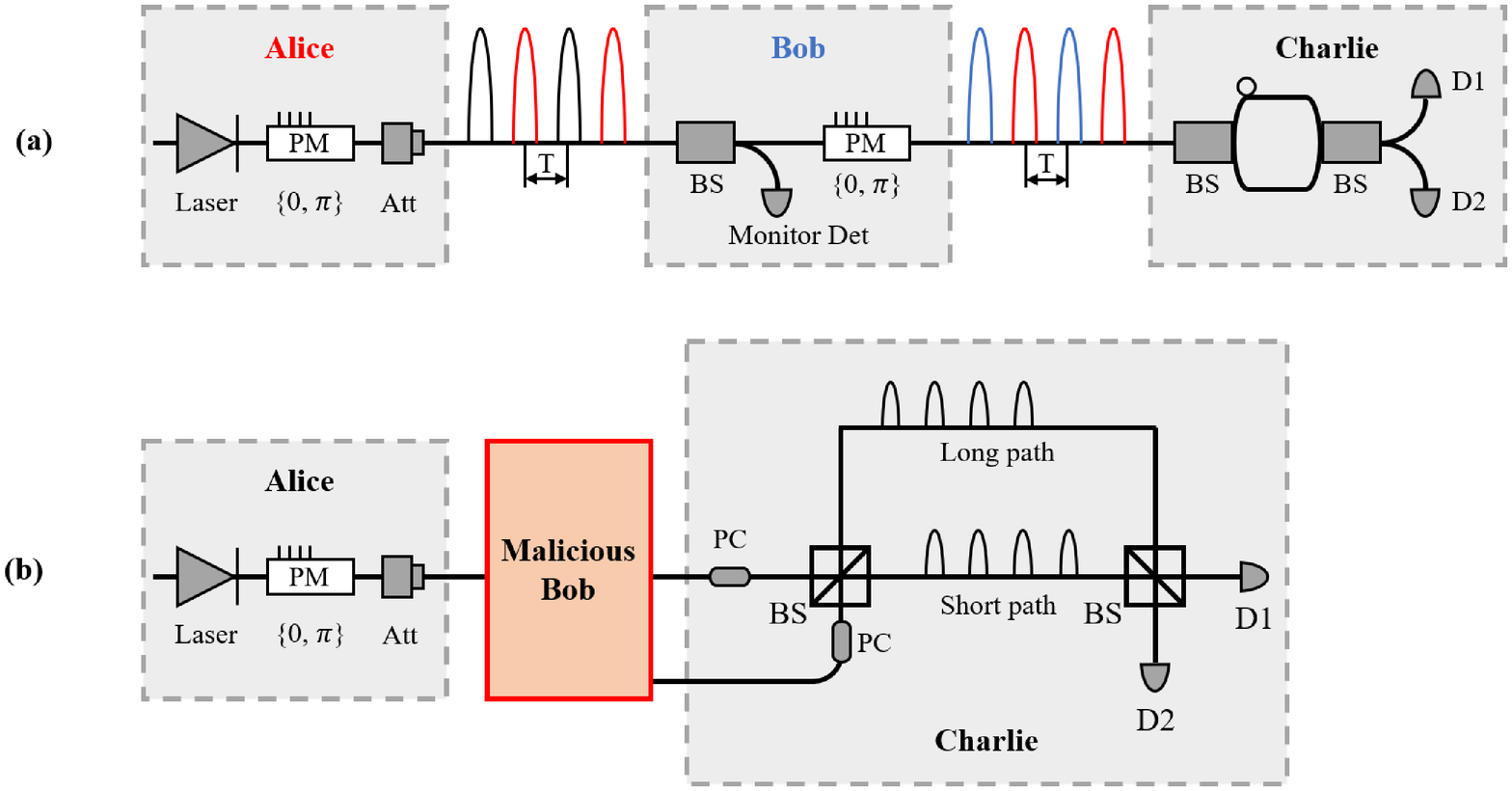}
	\caption{ {Internal eavesdropping of malicious Bob.} (a) With some special rules discussed in~\cite{gu2021differential}, we present the configuration of equivalence between differential phase shift QSS and our protocol. (b) We present the configuration of a general individual attack by malicious Bob.}
	\label{fig2}
\end{figure}

QSS protocols have to prohibit Alice or Bob from knowing Charlie's key by herself or himself. Firstly, we assume that Bob is the malicious one. In this case, we have equivalence configuration \cite{gu2021differential} shown in Fig.~\ref{fig2}a. Bob wants to know Charlie’s key by himself, and also needs to know Alice’s modulation phase to pass the test-bit checking among Alice, Bob, and Charlie after the creation of the raw key. A configuration for Bob to do so is shown in Fig.~\ref{fig2}b. He conducts general individual attacks as carried out by Eve in differential phase shift quantum key distribution \cite{PhysRevA.73.012344}, where the fraction Bob obtains about Charlie’s bits is $2\mu_a$. 
Similarly, when Alice is the malicious one, the probability that Alice knows Bob’s differential phase corresponding to Charlie’s bit is $2\mu_b$. From the above discussion, we denote $\mu_{\max}=\max\left\{\mu_a,\mu_b\right\}$ as the maximum intensity, then $2\mu_{\max}$ is the maximum ratio of information that a malicious one can obtain from the internal eavesdropping.

In conclusion, we find that the probability information leakage to Eve and malicious Bob (Alice) is $\mu_a(1-\eta_a)+\mu_b(1-\eta_b)$ and $2\mu_a{(2\mu_b)}$.
We discover that information leakage to external Eve is slightly lower than that to malicious Bob (Alice). For simplicity, we can just consider information leakage in our protocol to be $2\mu_{\max}$. 

	\section{Numerical Simulation}
		\subsection{Mathematical Calculation with Asymmetric Channels}
	Based on the asymmetric protocol description, Charlie generates their classical bits according to the phase differences between Alice and Bob in the same time slot. To obtain the secure key rate, we apply the same denotations in previous sections that Alice and Bob send pulses with intensities $\mu_a$, {$\mu_b$}, and {the} distance between Alice (Bob) and Charlie is $l_a (l_b)$. In our scheme, the channel transmittance between Alice (Bob) and Charlie is {$\eta_a=\eta_{d} \times 10^{-{\alpha l_a}/{10}} (\eta_b=\eta_{d} \times 10^{-{\alpha l_b}/{10}})$}, where $\eta_d$ is the detection efficiency of Charlie's detectors and $\alpha$ is the attenuation coefficient of the ultra-low fiber. In addition, let us suppose that $p_d$ is the dark count rate of one detector. For two detectors used by Charlie, we derive the total dark count rate as $2p_d$ and the error rate of background $e_0=\frac{1}{2}$.
	
	In Charlie's laboratory, after the BS (see Figure~\ref{fig1}), the optical intensities received by detector 1 and detector 2 are given by $D_1=(\frac{\sqrt{\mu_a\eta_a}}{2}+\frac{\sqrt{\mu_b\eta_b}}{2}\cos{\theta})^2$ and $D_2=(\frac{\sqrt{\mu_a\eta_a}}{2}-\frac{\sqrt{\mu_b\eta_b}}{2}\cos{\theta})^2$, where $\theta$ denotes the relative phase between Alice's and Bob's weak coherent states. In our asymmetric protocol, we have $\theta\in\left\{0,\pi\right\}$. Thus, the detection probability of each detector is: $Q_1=1-(1-p_d)e^{-D_1}$, $Q_2=1-(1-p_d)e^{-D_2}$.
	
	{The gain of the whole system for Charlie's detections can be calculated by $Q_\mu=Q_1(1-Q_2)+Q_2(1-Q_1)+Q_1Q_2$ and the error rate of the total gain can be derived by $	E_\mu Q_\mu=e_dQ_1(1-Q_2)+(1-e_d)Q_2(1-Q_1)+\frac{1}{2}Q_1Q_2$, where $e_d$ is the misalignment error rate of detectors.}

	\subsection{Finite-Key Analysis Method for Our Protocol}
	Considering the finite-key effects, let $ n_{\mu }= NQ_\mu$ be the observed number of bits, where $N$ is the number of optical pulses sent by Alice and Bob. By using the random sampling without replacement~\cite{yin2020tight}, one can calculate the upper bound of hypothetically observed error rate associated with $ E_{\mu }$ with a failure probability $\epsilon_{RS}$:
	\begin{equation}
		\overline{E_{\mu}}=E_{\mu}+\gamma(n_{\mu }-k,k,E_{\mu},\epsilon_{RS}),
	\end{equation}
	where {$k$} is the number of bits in the chosen time slots at Step 3 {(parameter estimation)}.

	In the following, we assume the protocol is $\epsilon$-secure~\cite{muller2009composability} where the maximum failure probability of
	practical protocol is $\epsilon$. According to universally composable security~\cite{959888},
	\begin{equation}	
		\epsilon=\epsilon_{RS}+\overline{\epsilon}+\epsilon_{EC}+\epsilon_{PA},
	\end{equation}
	where $\overline{\epsilon}$ represents the accuracy of estimating the smooth min-entropy. In addition, $\epsilon_{EC}$ corresponds to the probability that error correction fails and $\epsilon_{PA}$ is the probability that privacy amplification fails.	
	
	Then we obtain the key rate formula in finite-sized key region, which reads
	\begin{equation}
		\begin{aligned}
			R_{\rm QSS}=&Q_\mu[-(1-2\mu_{\max})\log_2(P_{\rm co})-f_eh(E_\mu)]\\
			&-\dfrac{7}{N}\sqrt{n_{\mu }\log_2{\dfrac{2}{	\overline{\epsilon}}}}-\dfrac{1}{N}\log_2{\dfrac{2}{\epsilon_{EC}}}-\dfrac{2}{N}\log_2{\dfrac{1}{\epsilon_{PA}}}.
		\end{aligned}\\
	\end{equation}

	Here, $f_e$ is the error correction efficiency and $h(x)=-x\log_{2}(x)-(1-x)\log_{2}(1-x)$ is Shannon entropy. $P_{\rm co}$ is the upper bound of collision probability when considering individual attacks, which can be concluded as $P_{\rm co} = 1-\overline{E_{\mu}}^2-{(1-6\overline{E_{\mu}})^2}/{2}$~\cite{lutkenhaus1999estimates}.
	
	\subsection{Results of Simulation}
\begin{figure}
		\includegraphics[width=8.6cm]{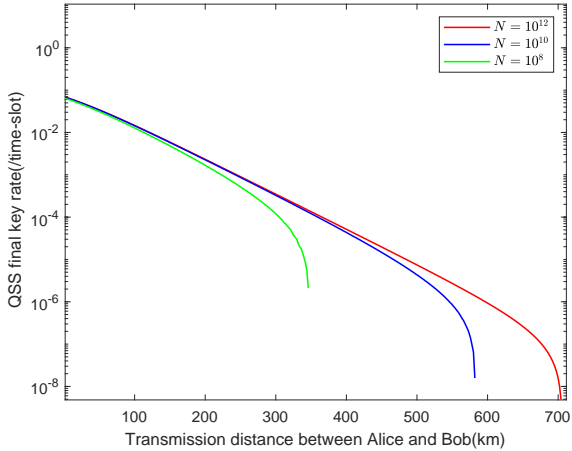}
		\caption{{The performance of our protocol under symmetric channels.} 
			Under the experimental parameters listed in Table.~\ref{tab2}, we simulate results in the case that $l_a=l_b$, where $N=10^{12},$ \mbox{$N=10^{10}$}, and $N=10^{8}$.}
		\label{fig3}
	\end{figure}

\begin{figure}
		\includegraphics[width=8.6cm]{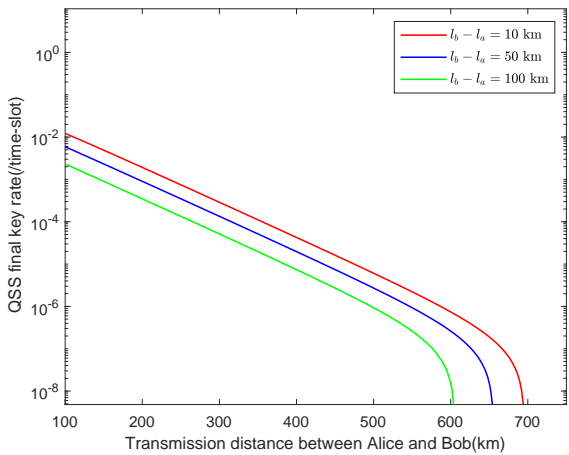}
		\caption{{The performance of our protocol under asymmetric channels.} 
			Under the experimental parameters listed in Table.~\ref{tab2}, we simulate results in the case that $N=10^{12}$, where $l_b-l_a=10, 50, 100$~km.}
		\label{fig4}
	\end{figure}

	We use the genetic algorithm to run the numerical simulations, and the key rate is optimized over the free parameters. Here, we set $\epsilon_{RS}=\overline{\epsilon}=\epsilon_{EC}=\epsilon_{PA}=10^{-10}$ and utilize experimental parameters listed in Table.~\ref{tab2}. Fig.~\ref{fig3} shows how the key rate varies with transmission distance between Alice and Bob when their channels have the same transmittance, where the total pulses are set as $N=10^{12},N=10^{10},N=10^{8}$, respectively.
	We can find that our protocol shows great robustness against finite-key effects.
	In \mbox{Fig.~\ref{fig4}}, we plot the results of our asymmetric protocol when the total pulses are set $N=10^{12}$, where the difference in length between Alice's channel and Bob's is fixed at 10 km, 50 km, and \mbox{100 km, respectively}.

	
\begin{figure}
		\includegraphics[width=8.6cm]{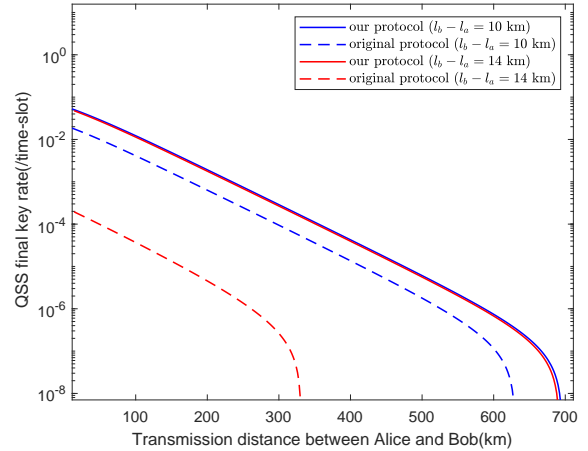}
		\caption{{Quantum secret sharing key rate vs transmission distance between Alice and Bob.} 
			Under the experimental parameters listed in Table.~\ref{tab2}, we compare the simulation results of the original TF-DPSQSS~\cite{gu2021differential} protocol and our asymmetric protocol
			with $N=10^{12}$, where the difference in length between Alice's and Bob's channels is fixed at 10 km, 14 km.}
		\label{fig5}
	\end{figure}

	When the original TF-DPSQSS~\cite{gu2021differential} protocol is applied to the asymmetric channels, a high system error rate will arise since different channel transmittances will lead to the poor performance of interference at the beam splitter. Fig.~\ref{fig5} presents numerical results of our asymmetric protocols and the original TF-DPSQSS~\cite{gu2021differential} protocol with the difference in length between two channels fixed at $10$ km and $14$ km. We can see clearly from \mbox{Fig.~\ref{fig5}} that our asymmetric protocol improves the secret key rate by two orders of magnitude when $l_b-l_a=14$ km. Moreover, no keys are obtained with difference fixed at 50, 100 km in the original protocol, whereas Fig.~\ref{fig4} shows that our protocol still provides a high key rate when the difference is large. It means that in the asymmetric channels, the performance of the asymmetric protocol is much better than that of the original protocol, especially when channels are extremely asymmetric.

	\begin{table}
		\centering
		\caption{Simulation parameters. $\eta_d$ and $p_d$ are the detection efficiency and dark count rate. $\alpha$ is the attenuation coefficient of the ultra-low fiber. $f_e$ is the error correction efficiency.}\label{tab2}
		\begin{tabular}{cccc}
			\hline
			\hline
			$\eta_d$ ~~~~~& $p_d$ ~~~~~& $\alpha$~~~~~ & $f_e$\\
			\hline
			$55\%$~~~~~& $10^{-8}$~~~~~ & $0.165$ ~~~~~ & $1.15$\\
			\hline
			\hline
		\end{tabular}
	\end{table}

	\section{Conclusion}
	In summary, we propose a differential phase shift quantum secret sharing protocol over asymmetric channels and give the security proof of our protocol against individual attacks. Moreover, we extend the asymptotic key rate of TF-DPSQSS~\cite{gu2021differential} to finite-key region. Through implementing free parameter optimization on the numerical simulations, we demonstrate that our asymmetric protocol can dramatically improve the key generation rate and the transmission distance compared with the original TF-DPSQSS~\cite{gu2021differential} protocol. As we have shown before, when the difference in length between Alice’s channel and Bob’s is fixed at 14 km, the key rate of the asymmetric protocol is two orders of magnitude higher than the original TF-DPSQSS~\cite{gu2021differential} protocol. Furthermore, our protocol obtains a high key rate when the difference is large. In addition, it is convenient and efficient to implement by allowing Alice and Bob to set asymmetric intensities independently, especially in a network setting. Due to the remarkable performance of our asymmetric protocol, it can be applied directly to the QSS experiments over asymmetric channels and represents a further step towards practical application of QSS.
	
	\section*{Acknowledgments}
	We acknowledge support from the National Natural Science Foundation of China (61801420); Key-Area Research and Development Program of Guangdong Province (2020B0303040001);  {Fundamental Research Funds for the Central Universities (020414380182, 020414380141).

	
	%

\end{document}